\documentclass[aps,prl,twocolumn,groupedaddress,showpacs,floatfix]{revtex4}
\def\bc{\begin{center}}
\def\ec{\end{center}}
\def\be{\begin{equation}}
\def\ee{\end{equation}}
\renewcommand{\vec}[1]{\mbox{\boldmath$#1$}}
\usepackage{longtable}
\usepackage{epsfig}
\begin{document}

\title{\bf Reconstructing the electron in a fractionalized quantum fluid} 
\author{Jainendra K. Jain and Michael R. Peterson} 
\affiliation{Department of Physics, 104 Davey Laboratory, The
Pennsylvania State University, Pennsylvania 16802} 
\date{\today}

\begin{abstract}
The low energy physics of the fractional Hall liquid is described in terms 
quasiparticles that are qualitatively distinct from electrons. 
We show, however, that a long-lived 
electron-like quasiparticle also exists in the excitation spectrum:
the state obtained by the application of an electron creation 
operator to a fractional quantum Hall ground state has a non-zero overlap 
with a complex, high energy bound state containing an odd number of 
composite-fermion quasiparticles.  The electron annihilation operator 
similarly couples to a bound complex of composite-fermion holes.  
We predict that these bound states 
can be observed through a conductance resonance in experiments involving 
a tunneling of an external electron into the fractional quantum Hall liquid.
A comment is made on the origin of the breakdown of the Fermi liquid 
paradigm in the fractional hall liquid.
\end{abstract}


\maketitle

The low energy excitations of an ordinary electron liquid resemble electrons.
During the past two decades, there has been much interest in systems where 
strong interactions may cause a breakdown of the Fermi liquid paradigm.
In the most dramatic instances, such a breakdown is signaled by the emergence 
of new quasiparticles that do not bear any resemblance to electrons, and 
even have quantum numbers which are a fraction of the electron
quantum numbers.    Is the electron irretrievably lost as a meaningful
entity in such a ``fractionalized" liquid?

In this Letter, we shall investigate this question in the context 
of the fractional quantum Hall liquid\cite{Tsui} (FQHL), formed when 
electrons are confined to two dimensions and subjected to a strong transverse 
magnetic field.  
The low-energy excitations of the FQHL carry a fractional 
charge\cite{Laughlin}.  The question of our interest is whether  
an integral number of such fractionally charged entities 
can combine to produce an electron.
The resolution requires a microscopic understanding of the strongly 
correlated FQHL state, which has been achieved in terms 
of composite fermions, bound states of 
electrons and an even number of quantized vortices\cite{Jain,Halperin}.  
The ground state of an incompressible FQHL is accurately described as a
state with an integral number of filled composite-fermion (CF) quasi-Landau 
levels (LL's), and its low-energy excitations are ``CF particles" (CFP's; 
composite fermions in otherwise empty CF-quasi-LL's) and ``CF  
holes" (CFH's; missing composite fermions in otherwise full CF-quasi-LL's).  
A CFP or a CFH has a fractional charge excess 
or deficiency associated with it\cite{Jain,Halperin,Jain2}, consistent with general 
principles that tell us that incompressibility at fractional 
fillings results in fractional charge\cite{Laughlin,Su}.  
It is obvious that an integral number of CFP's can have the same charge 
as an electron, but 
the key question is whether there exists a {\em long-lived} multi-CF bound 
complex that has a non-zero overlap with the ``electron quasiparticle," 
namely the excitation obtained by adding an electron to the ground state.  
Alternatively, can the electron be viewed as a stable bound state of CFP's? 
If so, what is that bound state?  How can it be observed?

We will use below the spherical 
geometry,~\cite{Haldane,JK} which takes $N$ electrons confined to 
the surface of a sphere and exposed to a radial magnetic field $B$ produced by a magnetic 
monopole of strength $Q$, which is restricted to be an integer or a half 
odd integer according to Dirac's quantization condition.  The single particle 
eigenstates of an electron in this geometry are a generalization of 
the usual spherical harmonics, called monopole harmonics, denoted by 
$Y_{Q,l,m}$, where $l=|Q|, |Q|+1, ...$ is the orbital angular momentum and 
$m = -l, -l+1, \cdots l$ is the $z$ component of the orbital angular momentum.
The different angular momentum shells are analogous to 
the Landau levels (LL's) of the planar geometry.   
The degeneracy of the lowest Landau level shell ($l=|Q|$) is $2|Q|+1$ 
(without counting spin), and increases 
by two units in each successive shell. 
It will be assumed below that the magnetic field is sufficiently strong that 
electrons are confined to the lowest Landau level 
and are fully spin polarized (the spin degree of freedom is frozen).
The only term remaining in the Hamiltonian is the Coulomb interaction, 
which determines the nature of the state in the lowest LL.

\begin{figure}
\center{\psfig{figure=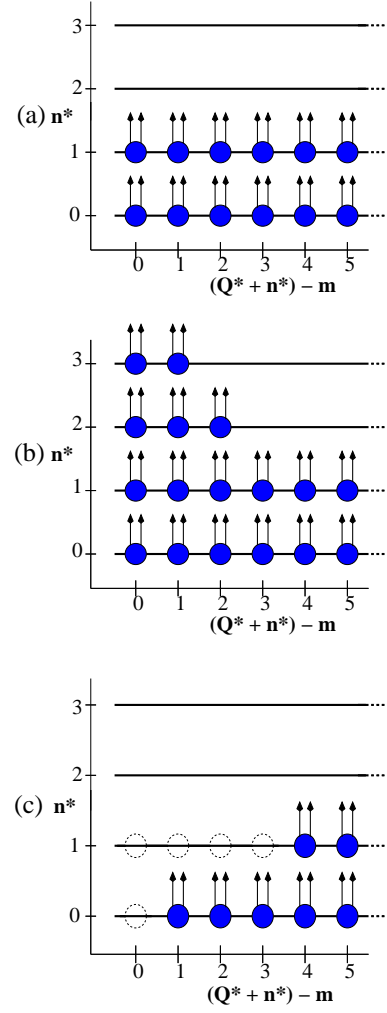,width=2.0in,angle=0}}
\caption{\label{fig1}
Schematic picture, at $\nu=2/5$, of the $(a)$ ground state, $(b)$ $\chi^+$, 
and $(c)$ $\chi^-$.
The composite fermions are depicted as dots (electrons) with two 
arrows (vortices) attached.  The vertical axis is the composite 
fermion quasi-LL index, $n^*$.  The horizontal axis shows $m$,  
the $z$-component of angular momentum of the composite fermion, 
translated by a suitable constant.  
For $\chi^-$ the CF holes are denoted by dashed circles (c). 
All higher $m$ states in the two lowest levels, not shown, are 
occupied.
}
\end{figure}

The CF theory postulates that electrons avoid one another most 
effectively by capturing an even number ($2p$)
of vortices to transform into composite fermions, which 
experience a reduced magnetic field,
produced by a monopole of strength $Q^*=Q-p(N-1)$.  
The wave functions $\chi$ for 
interacting electrons at $Q$ are constructed~\cite{Jain,JK} from 
the electron wave functions $\Phi^*$ at $Q^*$ according to 
\be
\chi={\cal P}_{LLL} \Phi_1^2\Phi^*
\ee
where ${\cal P}_{LLL}$ denotes projection into the lowest 
Landau level and $\Phi_1$ is the wave function of one filled 
Landau level.   An explicit, lowest-Landau-level-projected
form for $\chi$ can be obtained by methods
described in the literature~\cite{JK}.
We shall  consider $N$ particles at a monopole strength  
\begin{equation}
Q=(p+1/2n)N-(p+n/2)\;,
\label{QQ}
\end{equation}
where $p$ and $n$ are integers,
which is a finite size representation of the state at filling factor 
factor $\nu=\lim_{N\rightarrow\infty}{N}/{(2Q+1)}= n/(2pn+1)$.
It maps into a system of composite fermions at $Q^*={(N-n^2)}/{2n}$.
Here, $N$ composite fermions completely fill $n$ CF-quasi LL's,
which can be seen by noting that the total number of states in the 
lowest $n$ CF-quasi-LL's is $\sum_{j=0}^{n-1} (2Q^*+2j+1)=N$.
The ground state is a filled shell state, with 
total orbital angular momentum $L=0$, shown
schematically in Fig.~\ref{fig1} (a) for $\nu=2/5$.  Its wave function is 
$\chi_0={\cal P}_{LLL} \Phi_1^2\Phi_n$ 
where $\Phi_n$ is the Slater determinant wave function of $n$ filled Landau 
levels  at $Q^*$.  It is known~\cite{JK} to provide an excellent 
description of the exact ground state at $\nu=n/(2pn+1)$.

An electron quasiparticle (EQP) is created by adding an electron to 
the system in the lowest 
Landau level (while keeping $Q$ invariant), through 
application of the {\em projected} creation operator
\be
\bar{\Psi}^\dagger(\vec{\Omega})=\sum_m Y^*_{QQm}(\vec{\Omega})c^\dagger_{QQm}
\ee
where $\vec{\Omega}$ labels the position of the electron on the sphere and  
\begin{equation}
Y_{QQm}(\vec{\Omega})= \left[N_Q {{2Q}\choose {Q-m}}\right]^{1/2} v^{Q-m} u^{Q+m}
\end{equation}
with $u=\cos{\theta\over 2} e^{i\phi/2}$, $v=\sin{\theta\over 2} e^{-i\phi/2}$,
and $N_Q=(2Q+1)/4\pi$.
With no loss of generality, we add 
an electron at the North Pole ($\vec{\Omega}=\bar{\vec{\Omega}}$; $\bar{u}=1$, $\bar{v}=0$) 
by application of $\bar{\Psi}^\dagger(\bar{\vec{\Omega}})=\sqrt{N_Q} 
\;c^\dagger_{QQQ}$, onto the ground state wave function $|\chi_0\rangle$.  
Since we started out with a state with $L=0$, this 
gives us an EQP with $L=|M|=Q$, where $M$ is the z-component of the 
orbital angular momentum $L$.  The (unnormalized) wave function of the EQP is  
given by 
\be
\chi_{e}(\vec{\Omega}_1,\cdots,\vec{\Omega}_{N+1})=
A[Y_{QQQ}(\vec{\Omega}_{N+1}) \chi_0(\vec{\Omega}_1,\cdots,\vec{\Omega}_N)]
\ee
A ``hole quasiparticle" (HQP) at the North Pole is created 
similarly by application of the electron destruction operator 
$\bar{\Psi}(\bar{\vec{\Omega}})$.  The resulting (unnormalized) HQP wave function 
(with $L=|M|=Q$) is obtained by replacing one of the particle coordinates 
by $\bar{\vec{\Omega}}$
\be
\chi_{h}(\vec{\Omega}_1,\cdots,\vec{\Omega}_{N-1})=
\chi_0(\vec{\Omega}_1,\cdots,\vec{\Omega}_{N-1},\bar{\vec{\Omega}})\;.
\ee

Let us now analyze the problem of $N_{\pm}=N\pm1$ electrons at $Q$ according to the 
CF theory.  The new effective monopole strength is 
\be
Q^*_{\pm}=Q-p(N_{\pm}-1)=Q^*\mp p
\ee
Now, relative to $n$ filled CF-quasi-LL's, we have an excess of $2pn+1$ CFP's
or CFH's.  Thus, an EQP (a HQP) must be the bound state of 
$2pn+1$ CFP's (CFH's), consistent with the fact that a single CFP or CFH has 
a local charge of magnitude $e/(2pn+1)$.

For simplicity, we specialize to the $p=1$ sequence of fractions in what follows.
Let us consider the state at $Q^*_+$ with the lowest $n$ CF-quasi-LL's completely 
occupied and $2n+1$ CFP's remaining.  The lowest energy states are obtained by placing 
all of these CFP's in the lowest available CF-quasi-LL (which has 
single particle angular momentum $Q^*+n-1$), but all such configurations  
are strictly orthogonal to $\chi^+$ for symmetry reasons, because the largest possible 
total orbital angular momentum of these states, within the constraints of 
the Pauli principle, is less than the angular momentum $L=Q$ of the EQP.
It is therefore necessary for CFP's to occupy higher CF-quasi-LL's to make 
an EQP.  It can be easily verified that 
the lowest energy state with the desired 
angular momentum has $n+1$ and $n$ CFP's in the lowest two 
unoccupied CF-quasi-LL's, occupying the largest $m$ orbitals.
The corresponding wave function will be denoted by $\chi_+$.
Similarly, the lowest energy state with the quantum numbers matching those 
of $\chi_h$ is obtained by putting $n+2$ and $n-1$ CFH's in the top two 
occupied CF-quasi-LL's (for $n\geq 2$; for $n=1$ 
there are three CFH's in the only available CF-quasi-LL), again in the 
largest $m$ orbitals; its wave function will be denoted $\chi_-$.
Fig.~\ref{fig1} shows these bound complexes of CFP's and CFH's schematically 
for $\nu=2/5$ ($n=2$).  
Of course, many more states with $L=Q$ can be constructed, but they have 
at least one higher unit of CF-cyclotron energy compared to $\chi_+$ and $\chi_-$, 
and are therefore separated by a finite gap from $\chi_+$ and $\chi_-$ 
in the $L=|M|=Q$ subspace.  The wave function $\chi_+$ or $\chi_-$ 
can be constructed straightforwardly from the analogous wave function 
of $N_{\pm}$ electrons at $Q^*_{\pm}$.  
We have found that just like the ground state $\chi_0$, the CF complexes 
$\chi_{\pm}$ are also very accurate representations of the 
exact eigenfunctions of the lowest energy state in the $L=Q$ sector;
the overlap between $\chi_{\pm}$ and the corresponding exact states
are 0.98-0.99 for up to $N=9$.

\begin{figure}
\center{\psfig{figure=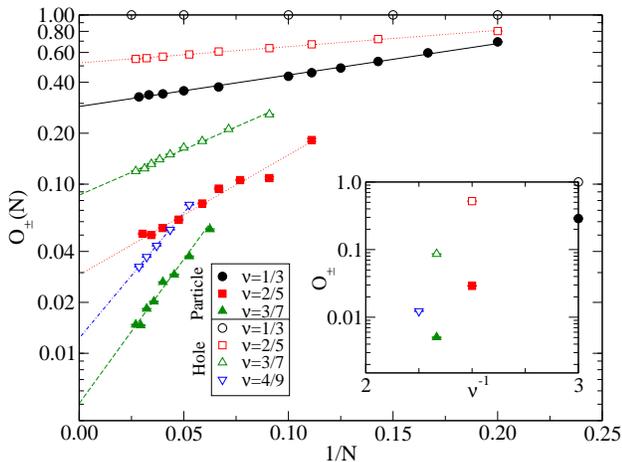,width=2.5in,angle=-90}}
\caption{\label{fig2}
The overlaps $O_{\pm}(N)$ for several filling factors 
of the form $\nu=n/(2n+1)$ as a function of the particle number $N$.
The inset shows the 
thermodynamic values of the overlaps $O_{\pm}$ as a function 
of $\nu^{-1}$; it decreases rapidly with $n$, but remains non-zero for 
all the fractions studied.
}
\end{figure}

To see how well $\chi_+$ and $\chi_-$ represent the EQP and HQP, we 
calculate the  overlaps 
$O_+=|\langle \chi_+|\chi_e\rangle|^2/(\langle \chi_+| \chi_+\rangle 
\langle \chi_e| \chi_e\rangle)$
and $O_-=|\langle \chi_-|\chi_h\rangle|^2/(\langle \chi_-| \chi_-\rangle 
\langle \chi_h| \chi_h\rangle)$, which are shown in Fig.~\ref{fig2} 
for several filling factors as a function of $1/N$.  For small $N$, 
$\chi_0$ and $\chi_{\pm}$ are explicitly very accurate, so  
$O_{\pm}$ are excellent approximations to the corresponding exact overlaps.  
That, we believe, remains true even in the thermodynamic limit, although 
there, strictly speaking, our calculation represents a prediction 
to be tested experimentally (see below).
Our principal finding is that the 
overlaps approach a non-vanishing value in the limit $N^{-1}\rightarrow 0$,
demonstrating that a rather complicated 
bound complex of CFP's (CFH's) has in it a non-zero content 
of an EQP (a HQP).  The thermodynamic overlap decays rapidly with  
$n$ along the sequence $\nu=n/(2n+1)$, as expected from the increasing complexity 
of the multi-CF bound state.
It has been noted earlier~\cite{Rezayi} that for $\nu=1/(2p+1)$, where the wave functions 
for the CF ground state and the CFH are identical to those written earlier 
by Laughlin,~\cite{Laughlin} we have $O_-=1$.

The existence of an electron-like multi-CF bound complex 
is not merely a theoretical curiosity but 
has testable consequences for experiments that 
involve tunneling of an electron into an incompressible FQHL~\cite{Ashoori,Eisenstein}.
An example is vertical interlayer transport in  
a bilayer system~\cite{Eisenstein} in the weak coupling limit 
(when  the nature of the state in either layer is not
affected by its proximity to the other layer), with each layer being 
at $\nu=n/(2pn+1)$.  For the tunneling Hamiltonian $T\int d^2\vec{r} 
\bar{\Psi}_1^\dagger(\vec{r})\bar{\Psi}_2(\vec{r})+H.c.$ (1 and 2 label 
the two layers; the tunneling amplitude $T$ is taken to 
be energy indepenent in the relevant energy range), the tunneling current at 
zero temperature is proportional to~\cite{Mahan} 
\be
I(eV)\propto \int_0^{eV} dE \bar{\rho}^{(+)}(E) \bar{\rho}^{(-)}(eV-E)
\ee
where $eV$ is the bias voltage and 
$\bar{\rho}^{(+)}(E)$ and $\bar{\rho}^{(-)}(E)$ are the positive and 
negative frequency parts of the projected electron spectral function:
\begin{eqnarray}
\bar{\rho}^{(\pm)}(E)&=&\sum_m \frac{|\langle m| \bar{\Psi}^{(\pm)}(\vec{0})|0\rangle|^2 
}{\langle m|m\rangle \langle 0|0\rangle}
\delta(E-E^{N\pm1}_m+E_0)\nonumber \\
&=&\nu_{\pm} N_Q  \sum_m  O_{\pm}^{(m)}
\delta(E-E^{N\pm1}_m+E_0)
\end{eqnarray}
Here $|0\rangle$ is the exact ground state of $N$ particles with 
eigenenergy $E_0$, and $|m\rangle$ labels all 
exact eigenstates (only those with $L=Q$ are relevant here) of 
the $N\pm 1$ particle systems with eigenergy $E^{N\pm 1}_m$.
The symbols are defined as $\nu_+=\nu$, $\nu_-=1-\nu$, $\bar{\Psi}^{(+)}=
\bar{\Psi}^\dagger$, $\bar{\Psi}^{(-)}= \bar{\Psi}$, and 
$$O_{\pm}^{(m)}\equiv\frac{|\langle m | \bar{\Psi}^{(\pm)}(\vec{0})|0\rangle|^2}
{\langle m|m\rangle \langle 0|[\bar{\Psi}^{(\pm)}(\vec{0})]^\dagger\bar{\Psi}^{(\pm)}(\vec{0})|0\rangle}
\; .$$
We have also made use of
$$\langle 0|\bar{\Psi}^\dagger(\vec{0}) \bar{\Psi}(\vec{0}) |0\rangle
= N_Q \langle 0|c_{QQQ}^\dagger c_{QQQ} |0\rangle
= N_Q \nu  \langle 0|0\rangle.
$$
We approximate the exact oscillator  
strengths $O_{\pm}^{(0)}$ for the lowest energy 
states at $L=Q$ by the CF overlaps $O_{\pm}$ calculated earlier.
Non-zero values for $O_+$ and $O_-$ in the 
thermodynamic limit imply delta function peaks 
at $E=E_{\pm}-E_0$ in the electron spectral function, which in turn 
produces a sharp peak in the conductance at a voltage $V$ given by $eV=E_++E_--2E_0$.
For $\nu=1/3$, $2/5$, and $3/7$, we have estimated $E_++E_--2E_0$ to be 
$\sim 0.5 e^2/\epsilon \ell$ by an extrapolation to $N^{-1}\rightarrow 0$,
where $\epsilon$ is the dielectric constant of the host semiconductor and
$\ell=\sqrt{\hbar c/eB}$ is the magnetic length.  
The ``coherent" peak is expected to be followed by a broad ``incoherent" peak where
the tunneling electron couples into a quasi-continuum of 
higher energy excited states.

The delta function peaks in the projected 
spectral function indicate that the 
electron and hole quasiparticles are long lived.  Lower energy states 
do exist, but the EQP's or HQP's cannot decay into them 
on account of angular momentum conservation.   
In practice, disorder, left out in the above analysis,
will impart a non-zero width to the quasiparticle peak; 
it is not possible at present to estimate the broadening 
for lack of a quantitative understanding of the effect of disorder.

Tunneling experiments in bilayer systems have been performed in the 
past~\cite{Eisenstein}.  The lack of tunneling at small voltages and the 
presence of a broad conductance peak at a finite voltage has been well 
understood~\cite{Johansson,Hatsugai,He,Pikus}, attributed
to a Coulomb gap resulting from the
strongly correlated nature of the electron system in either layer.
It is natural to identify the observed peak as arising from the 
incoherent part of the spectral function, given its lack of 
sensitivity to the details of the correlations:  
the peak is independent of the filling factor; it occurs  
for both incompressible and compressible ground states~\cite{Hatsugai,He}; 
one may even model the liquid ground state
as a Wigner crystal to understand its origin~\cite{Johansson}; and the
energy gap can be estimated from simple classical electrostatic 
considerations~\cite{He,Pikus}.  The coherent peak discussed in this work, 
on the other hand, is crucially dependent
on the physics of the fractional quantum Hall effect and occurs only
for incompressible states, with a strongly $\nu$ dependent
oscillator strength.  It is noteworthy  
that the theory of Conti and Vignale~\cite{Conti}, which  
models the bulk as a continuous elastic medium and the collective 
excitations as bosons, predicts a coherent peak in the spectral function 
for $\nu=1/3$, with additional structure in the 
incoherent peak.

An observation of a sharp coherent resonance in tunneling experiments will  
provide new insights into the nature of the FQHL.  
There can be many reasons for its absence in the earlier experiments.
While the tunneling from one layer to another takes place predominantly in 
the bulk, the current is being injected and collected at a lead connected to an 
edge of the sample.  The passage of tunnel current from the bulk to the lead 
not only exaggerates the effect of disorder, but also 
requires a reasonably high temperature (current flow in the bulk is 
exponentially suppressed at low temperatures).  It is possible that the 
combined broadening due to 
temperature and disorder has suppressed the coherent peak in the 
previous experiments.  Reducing disorder 
and the sample area might help, as might a tri-layer FL-FQHL-FL (FL $=$ Fermi liquid) 
geometry, in which the tunneling electron passes right through the FQHL layer.

Before ending, we make an  
observation on the above results vis-\`{a}-vis the general question 
of the origin of ``non-Fermi liquid" behavior.  
The fundamental tenet of the Fermi liquid theory is that electron-like
quasiparticle is long lived, i.e., there 
is a delta function peak (broadened into a Lorenzian shape by disorder)
in the electron spectral function, with weight $Z$, called the renormalization
factor.  One possible mechanism for the breakdown of the Fermi liquid
is the vanishing of $Z$.  That is the case in
a well understood example of a non-Fermi liquid, namely the
Luttinger liquid in one dimension.  The FQHL provides a different
paradigm.  Here, the electron like quasiparticle remains well defined, with a 
non-zero $Z$, but the Fermi liquid description breaks down 
because other excitations appear at {\em lower} energies, described in terms of 
new elementary quasiparticles that are qualitatively distinct 
from electrons.

In summary, we have shown that an electron-like quasiparticle exists  
in the fractional quantum Hall liquid, but has a strikingly complicated structure.
It is a complex ``atom" of an odd number of composite fermions, 
which are themselves collective bound states of electrons and quantized 
vortices of the wave function.
A direct signature of this bound complex should appear as a conductance resonance in 
vertical tunneling transport in bilayer systems.

Partial support of this research by the National Science Foundation under Grants
No. DMR-0240458 and DGE-9987589 (IGERT) is gratefully acknowledged.

\end{document}